\documentclass{aastex61}

\newcommand\aastex{AAS\TeX}

\received{June 1, 2019}
\revised{January 10, 2019}
\accepted{\today}
\submitjournal{ApJL}

\usepackage{color}
\usepackage{epstopdf}
\usepackage{graphicx}
\usepackage[FIGTOPCAP]{subfigure}
\usepackage{amsmath}

\UseRawInputEncoding

\begin{document}

\begin{center}
Using {\tt\string natbib} with \aastex
\end{center}

\title{Observations of Forbush Decreases of cosmic ray electrons and positrons with the Dark Matter Particle Explorer}

\correspondingauthor{Jing~Jing~Zang}
\email{zangjingjing@lyu.edu.cn}

\author{Francesca~Alemanno}
\affiliation{Gran Sasso Science Institute (GSSI), Via Iacobucci 2, I-67100 L¡¯Aquila, Italy}
\affiliation{Istituto Nazionale di Fisica Nucleare (INFN) - Laboratori Nazionali del Gran Sasso, I-67100 Assergi, L¡¯Aquila, Italy}

\author{Qi~An}
\affiliation{State Key Laboratory of Particle Detection and Electronics, University of Science and Technology of China, Hefei 230026, China}
\affiliation{Department of Modern Physics, University of Science and Technology of China, Hefei 230026, China}

\author{Philipp~Azzarello}
\affiliation{Department of Nuclear and Particle Physics, University of Geneva, CH-1211, Switzerland}

\author{Felicia~Carla~Tiziana~Barbato}
\affiliation{Gran Sasso Science Institute (GSSI), Via Iacobucci 2, I-67100 L¡¯Aquila, Italy}
\affiliation{Istituto Nazionale di Fisica Nucleare (INFN) - Laboratori Nazionali del Gran Sasso, I-67100 Assergi, L¡¯Aquila, Italy}

\author{Paolo~Bernardini}
\affiliation{Dipartimento di Matematica e Fisica E. De Giorgi, Universit\`a del Salento, I-73100, Lecce, Italy}
\affiliation{Istituto Nazionale di Fisica Nucleare (INFN) - Sezione di Lecce, I-73100, Lecce, Italy}

\author{XiaoJun~Bi}
\affiliation{University of Chinese Academy of Sciences, Yuquan Road 19A, Beijing 100049, China}
\affiliation{Particle Astrophysics Division, Institute of High Energy Physics, Chinese Academy of Sciences, Beijing 100049, China}

\author{MingSheng~Cai}
\affiliation{Key Laboratory of Dark Matter and Space Astronomy, Purple Mountain Observatory, Chinese Academy of Sciences, Nanjing 210023, China}
\affiliation{School of Astronomy and Space Science, University of Science and Technology of China, Hefei 230026, China}

\author{Elisabetta~Casilli}
\affiliation{Dipartimento di Matematica e Fisica E. De Giorgi, Universit\`a del Salento, I-73100, Lecce, Italy}
\affiliation{Istituto Nazionale di Fisica Nucleare (INFN) - Sezione di Lecce, I-73100, Lecce, Italy}

\author{Enrico~Catanzani}
\affiliation{Istituto Nazionale di Fisica Nucleare (INFN) - Sezione di Perugia, I-06123 Perugia, Italy}

\author{Jin~Chang}
\affiliation{Key Laboratory of Dark Matter and Space Astronomy, Purple Mountain Observatory, Chinese Academy of Sciences, Nanjing 210023, China}
\affiliation{School of Astronomy and Space Science, University of Science and Technology of China, Hefei 230026, China}

% ...

\collaboration{146}{(DAMPE Collaboration)}

\author{Wei~Chen}
\affiliation{Key Laboratory of Dark Matter and Space Astronomy, Purple Mountain Observatory, Chinese Academy of Sciences, Nanjing 210023, China}
\author{Li~Feng}
\affiliation{Key Laboratory of Dark Matter and Space Astronomy, Purple Mountain Observatory, Chinese Academy of Sciences, Nanjing 210023, China}
\author{Xi~Luo}
\affiliation{Shandong Institute of Advanced Technology, 250100 Jinan, China}
\author{ChengRui~Zhu}
\affiliation{Key Laboratory of Dark Matter and Space Astronomy, Purple Mountain Observatory, Chinese Academy of Sciences, Nanjing 210023, China}

\begin{abstract}
The Forbush Decrease (FD) represents the rapid decrease of the intensities
of charged particles accompanied with the coronal mass ejections
(CMEs) or high-speed streams from coronal holes. It has been mainly
explored with ground-based neutron monitors network which indirectly measure
the integrated intensities
of all species of cosmic rays by counting secondary neutrons produced from interaction between atmosphere atoms and cosmic rays. The space-based experiments can resolve the species of particles but the energy ranges are limited by the relative small acceptances except for the most abundant particles like protons and helium. Therefore, the FD of cosmic ray electrons and positrons have just been investigated by the PAMELA experiment in the low energy range ($<5$ GeV) with limited statistics. In this paper, we study the FD event occurred in September,
2017, with the electron and positron data recorded by the Dark Matter
Particle Explorer. The evolution of the FDs from 2 GeV to 20
GeV with a time resolution of 6 hours are given. We observe two solar energetic particle events in the time profile of the intensity of cosmic rays, the earlier and weak one has not been shown in the neutron monitor data. Furthermore, both
the amplitude and recovery time of fluxes of electrons and positrons
show clear energy-dependence, which is important in probing the
disturbances of the interplanetary environment by the coronal mass ejections.
\end{abstract}


\section{Introduction} \label{sec:intro}

Charged solar wind particles and the associated magnetic fields affect
the transportation of Galactic cosmic rays (GCRs) in the solar system.
Violent solar activities like coronal mass ejections (CMEs) generate
high-intensity particle flows moving at super-Aflven speed, which enhance
the local interplanetary magnetic field significantly. Such temporarily
enhanced magnetic fields would block the propagation of GCRs,
what contributes to sharp decreases of the GCRs fluxes,
known as Forbush decreases \citep[FDs;][]{Forbush1937, Hess1937}.
The Forbush decrease is a universal phenomenon within the heliosphere,
which was also observed at other planets, e.g., the Mars
\citep{FDMars2018} and the interplanetary space far away from
the Sun (e.g., by Voyager 2 \citep{FDvoyageII2015}).

Precise measurements of FDs will enable us to diagnose the propagation
of GCRs and their interplay with complex environment in the heliosphere.
FDs of GCRs have been extensively measured for decades with worldwide
ground-based neutron monitors (NMs) located at regions with different
geomagnetic cutoff rigidities. Compared with direct detection
experiments, the NMs can only reflect an integral variation of the
incident particle intensities, with much information of the compositions
and precise energy-dependencies losing.
In the 1960s, some balloon experiments \citep{1961JGR....66.3950M} measured
FDs of electrons and protons at MeV energies. However, the relatively poor
energy resolution, bad particle identification, and limited statistics
hindered a good understanding of FDs of various particle species.
Recently, the PAMELA experiment simultaneously measured FDs of
protons, helium nuclei, and electrons for the first time, and found
important differences of the properties of different particle species
\citep{PAMELAFD2018}. Specifically, the recovery time for electrons
is faster than that for protons and helium nuclei, which may be
interpreted as a charge-sign-dependent drift pattern between electrons
and nuclei \citep{PAMELAFD2018}. However, due to its relatively small geometry
factor, the data statistics is limited (particularly for electrons),
and the energy-dependence of the FD characteristics remains somehow
ambiguous.

The Dark Matter Particle Explorer (DAMPE) is a space-borne detector
for observations of cosmic ray electrons and positrons (CREs), nuclei,
and $\gamma$-ray photons \citep{TheDAMPE2017, Chang2017b, Proton2019}.
The DAMPE is optimized for high-energy-resolution measurements of CREs.
Compared with other particle detectors in orbit, the DAMPE has two
unique advantages in the measurements of time-variations of CREs fluxes,
1) the inclination angle of the orbit is 97 degrees \citep{TheDAMPE2017},
and thus DAMPE can reach the Earth's polar regions where GCRs are weakly
affected by the geomagnetic rigidity cutoff, and 2) the relatively
large effective geometry acceptance ($\sim0.35$~m$^{2}$sr) enables a
more detailed study of the fine time structures of FDs.
Benefited from these two facts, in this work we present the observations
of FDs of CREs by DAMPE with unprecedented accuracy and give solid evidence
of energy-dependences for both the FD amplitude and the recovery time.

This paper mainly focuses on a strong FD event occurred in September,
2017 when two shock-associated interplanetary coronal mass ejections
(ICMEs) and a ground level enhancement (GLE) were also detected by
on-ground NMs. The solar flares in September 2017 are very famous events. Based on the NM data, \citet{Badruddin2019} presented
3-4 hours time-lagged correlation between cosmic ray intensities and
the geomagnetic activity indices. \citet{Hubert2019} presented cosmic ray-induced neutron spectra during active solar event leading to changes in the local cosmic ray spectrum (Forbush decreases and a GLE). Meanwhile, by calculating the GCR spectral index of FDs, several theoretical yield functions were verified in \citet{Livada2020} and
references therein. \citet{Chertok2018SolarEF} analyzed space weather disturbance and successfully estimated the scale of FD and geomagnetic storm.



\section{DAMPE detector}\label{sec:DAMPEDetector}
The DAMPE satellite operates in a sun-synchronous orbit at 500 km
altitude with a period of 5673 seconds and an incline angle of 97 degrees that enables the satellite to travel the area from 83 degrees north to 83 degrees south.  The detector system consists
of four sub-detectors \citep{TheDAMPE2017}. At the top is a plastic scintillator array detector
(PSD) which measures the ionization energy loss of charged particles and
also acts as an anti-coincidence veto detector for identification of
photon candidates \citep{2017APh....94....1Y}. The instrument mounted
below PSD is a silicon tungsten tracker (STK) equipping with 12 layers
of $121~\mu$m wide silicon strip sensors \citep{2016NIMPA.831..378A}.
STK is designed to measure the trajectory and charge of a charged
particle and to convert $\gamma$-rays into $e^+e^-$ pairs. The core
detector of DAMPE is a Bismuth Germanate (BGO) calorimeter which has 32 radiation
lengths thickness of BGO crystals, and is to measure the energy and
direction of a particle, to distinguish hadronic and electromagnetic
particle species, and also to provide the trigger for the Data
Acquisition (DAQ) system \citep{2015NIMPA.780...21Z}. The dense and
thick materials of the BGO provides an excellent energy resolution for
CREs, which is about 5\% (2\%) at $\sim 2$ (20) GeV \citep{BGOBT2016}.
At the bottom is a neutron detector (NUD) which is used to improve the
capability of hadron-nuclei separation, since much more neutrons are
generated in hadronic shower. The on-orbit performance of each sub-detector
can be found elsewhere \citep{2018NIMPA.893...43T,Ambrosi2019,
2019RAA....19...82M,TYKHONOV2019309,Huang2020}.

\section{Data Analysis}
We describe the time variation of the GCRs due to the CME event at two
levels, the counting rate from the T0 counts (a category of data acquisition trigger with very loose threshold, see below for details) and the absolute fluxes of CREs. The T0 counting rate measures the overall intensity of all species of GCRs above 100MeV with averaged 48 minutes(half orbit period) time resolution, that enabled DAMPE to discover fine structures of time profile of GCRs intensity. Meanwhile, the absolute fluxes variations of CREs in dozens energy intervals are simultaneously measured for the purpose of investigating the energy dependence of the characteristics of FDs.

\subsection{T0 counting rate}

\begin{figure}[htb]
  \centering
  \plotone{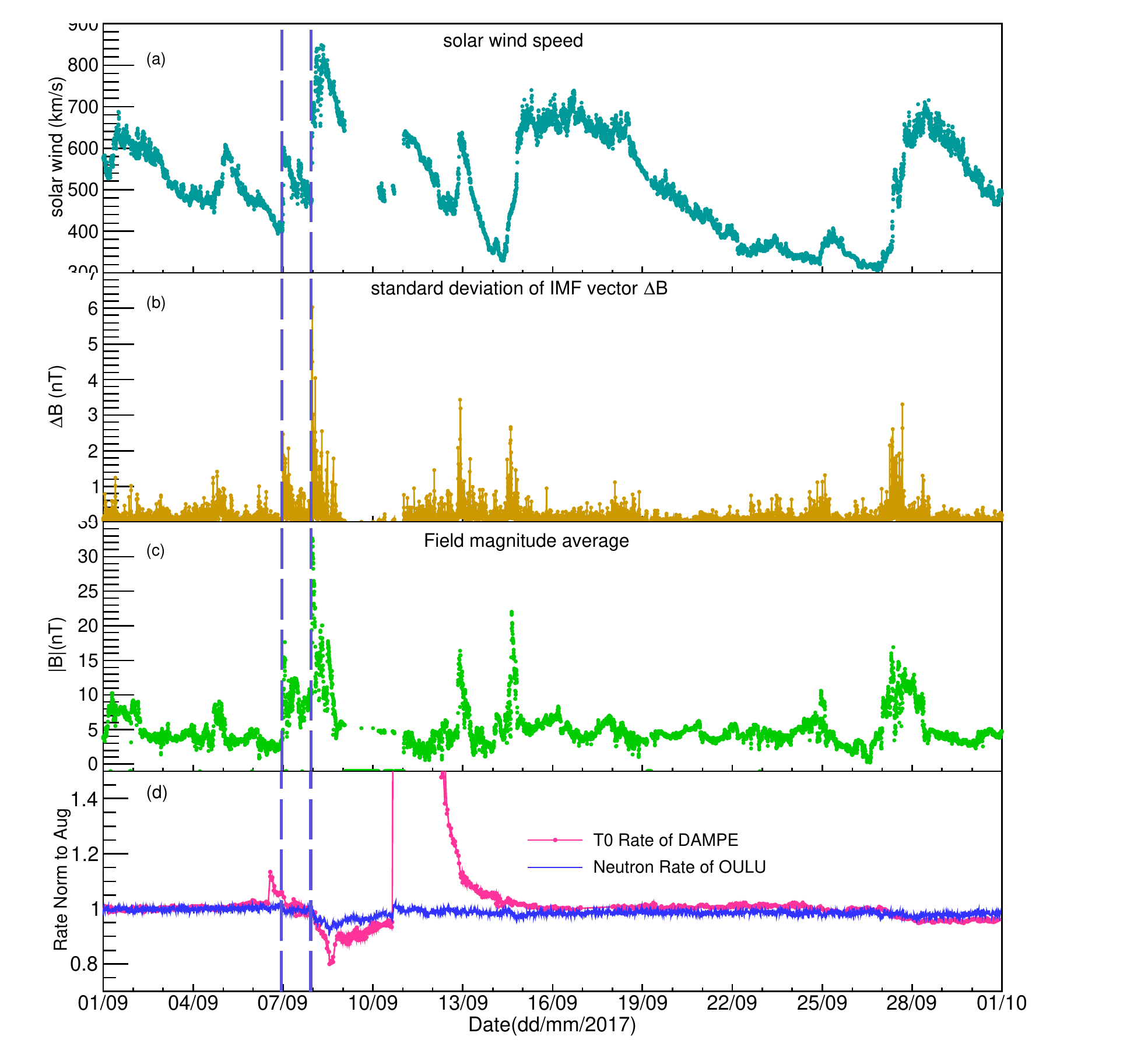}
  \caption{The 5-min averaged time profiles of the solar wind speed (panel (a)), the standard deviation of the IMF vector (panel (b)), the averaged strength of the IMF at 1 AU (panel (c)). Data in panel(a)(b)(c) are all from the OMNI database(\url{https://omniweb.gsfc.nasa.gov/form/omni\_min.html}). Panel(d) shows the 47-min averaged temporal variation of T0 counting rate of DAMPE for $4.76<L<5.57$ (Rate is normalized to averaged value in August 2017). Also shown in the panel (d) is the OULU NM 10-min averaged relative counting rate with respect to averaged value in August 2017 (data from \url{http://www01.nmdb.eu/nest/}). The two vertical red lines indicate the arrival times of the interplanetary shocks.\label{fig:T0}}

\end{figure}

During the on-orbit operation of DAMPE, the housekeeping system records
GCRs trigger counts every 4 seconds (named T0 trigger count) for the purpose
of monitoring the health of detectors. T0 is a low threshold hit signal acting as the timing reference to decide to start an event data taking. Based on Geant4 simulation, GCRs with kinematic energy above 100MeV can satisfy the T0 trigger logic. Due to T0 does not record information of particle species, its counting rate only
reflects the overall intensity of all kinds of GCRs above 100MeV. On DAMPE's orbit, the T0 counting rate changes
from several hundred Hz near the equator to several thousand Hz at poles.
We would like to compare the DAMPE T0 counting rate with the NM data.
Since the NMs are fixed on the surface of the Earth, and DAMPE keeps
moving along its orbit, we select the T0 counting rate in a narrow
region with specific MacIlwain L-parameter values \citep{Lparameter1994} in there geomagnetic cutoff rigidity is assumed to be a constant.
For the OULU NM we adopt in the comparison, we find that for
$4.76<L<5.57$ the averaged cutoff rigidity is $R_c\approx0.81$~GV which is similar
with that of the OULU station. The panel (d) of Figure \ref{fig:T0}
shows the time evolutions (normalized to the average values in August,
2017) of the DAMPE T0 rate (red) and the OULU NM rate (blue) in
September, 2017. A 27 days long-term shallow change of the T0 rate has been
subtracted through dividing the rate by a  smooth polynomial fitting
form to the T0 counts from August 23 to September 24 after abandoning the
FD time range. Also shown in Figure \ref{fig:T0} are 5-min averaged time profiles of the solar wind speed
(panel (a)), the standard deviation of the interplanetary
magnetic field (IMF) vector (panel (b)), and the
average strength of the IMF (panel (c)).
Two vertical lines label the arrival times\footnote{Data from the CfA
Interplanetary Shock Database: \url{www.caf.harvard.edu}} of two interplanetary
shocks, at 23:02 UT, September 6, 2017 and 22:28 UT, September 7, 2018.

From the T0 counting rate, we can clearly identify two solar
energetic particle (SEP) events. The first one with smaller amplitude
reached the Earth $10-11$ hours before the interplanetary shock wave
and lasted for about 20 hours. However, the NM data (from OULU and others)
did not show such a rising for this event (see panel (d) of
Figure~\ref{fig:T0}). The reason of the difference between NMs and the
direct measurements is unclear yet. To a certain extent, DAMPE is not
blocked by the atmosphere and is more sensitive to low-energy particles
than NM stations. The other possible reason is the anisotropy of SEPs
which may lead to missing of the detection by NMs.
We also investigate the time profiles of the T0 counts at high latitudes
(with $L>4.11$) in both southern and northern hemispheres, and such
intensity enhancements are both observed.
About 89 hours after the arrival of the second shock (during the recovery
phase of the FD), the T0 counting rate further shows a sudden and big
jump (increased by 2 orders of magnitude), which indicates the arrival
of the second SEP event. At that moment the OULU NM also observed a GLE
event with several percents increase of the counting rate. The big
difference of the SEP amplitudes between DAMPE and OULU is again a
reflection that NMs are only sensitive to high enough energy particles.

During SEPs passbying the Earth (UT 0906-23:02 to 0910-16:03) the fluxes
of energetic particles are so high that the detector cannot operate in
the normal status any more with significant and random shifts of the
pedestals of the electronics. Thus the science data taken in this period
has been excluded in the current analysis.

\subsection{CRE fluxes}

For the purpose of this study, about 60 days of the DAMPE data, in total
300 million events, recorded in August and September 2017 are analyzed.
To make sure the detector operates with good condition,
the data recorded when DAMPE passes through the South Atlantic Anomaly
(SAA) region are excluded.

We apply a series of selection criteria to filter a clean sample of CREs.
The purpose is to suppress the background cosmic rays as much as possible
while keeping a sufficiently high CRE efficiency. The first step of the
pre-selections is to require that the BGO bar with the maximum deposited
energy in each layer is not at the edge of the BGO calorimeter, which
can make sure that most of the shower energies are recorded by the
calorimeter and a good energy resolution is guaranteed. Then a very
similar track selection and evaluation strategy as described in
\cite{photon2017} is adopted to identify the tracks of events.
Specifically, the STK track should be a long track crossing all layers of STK and matches the BGO track, and the CRE should deposit most energy within a 5mm cylinder around the track.
Once the track is determined, the charge of the particle can be
obtained via the charge reconstruction algorithm \citep{charge2019}.
Here we require that the charge number measured by PSD should be less
than 1.7, which can reject $Z\ge2$ nuclei up to a level of $99\%$.
Finally, to eliminate secondary particles generated in the atmosphere,
the minimum energy at each geomagnetic latitude is required to exceed 1.2
times of the vertical rigidity cutoff \citep[VRC;][]{VRC2005}.

The pre-selections can already strongly exclude heavy nuclei from the
sample. However, there is still heavy contamination from protons and also
helium nuclei. Based on the shower development in the calorimeter, a
particle identification (PID) variable is constructed.
$${\rm PID} = F(E)[\log({\rm RMS}_r)\sin\theta+\log({\rm RMS}_l)\cos\theta]$$
where, $F(E)$ is energy decoupling polynomial, $\theta$ is coordinate
rotation angle, ${\rm RMS}_r$ and ${\rm RMS}_l$ are functions describing
radial shower shape and longitudinal shower shape respectively.

\begin{figure}[htb]
\centering
\includegraphics[width=.45\textwidth]{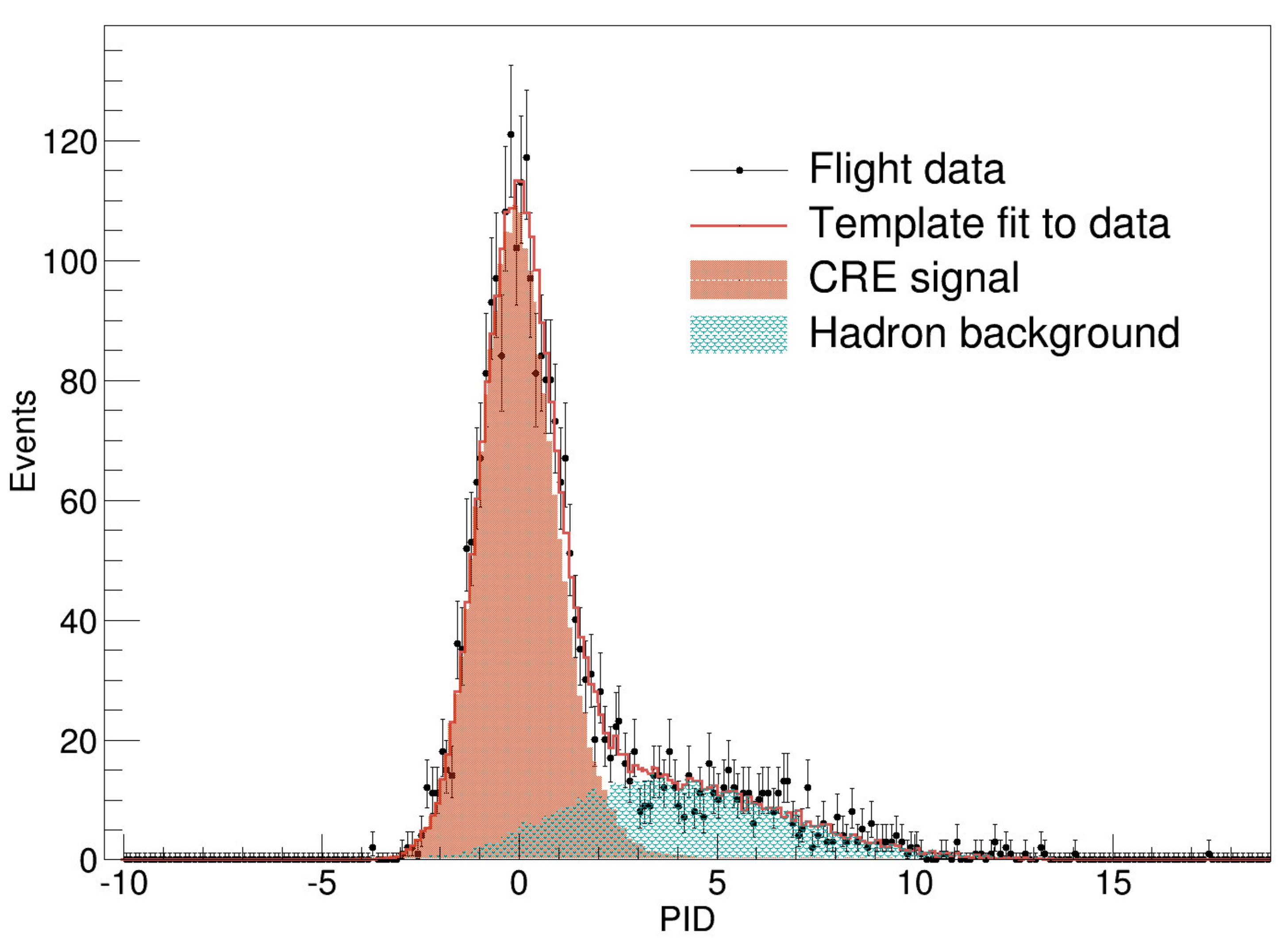}
\caption{Distributions of the PID variable for deposited energy bin
$4.83$~GeV$<E<5.25$~GeV. The best-fit MC templates are also shown for
comparison. Setting the PID selection value of $<2$ results
in a background contamination of $\sim8\%$.}
\label{fig:PID}
\end{figure}

The background contamination can then be estimated through a fitting
with Monte Carlo (MC) templates in each energy bin, as illustrated in
Figure \ref{fig:PID}. The background contamination is estimated to be
$2\%\sim8\%$ for events with deposited energies between 2 GeV and 20 GeV.
Below 2 GeV, the granularity of the calorimeter does not allow us to
efficiently identify CREs from nuclei. Since the low energy particles
are modulated by solar winds, their fluxes show slight and continuous
variations with time. We thus estimate the background in each time bin.
It is found that the hadronic background varies by $\sim6\%$ at 2 GeV
and $\sim1\%$ at 20 GeV.
After the pre-selections and the PID procedure, the CRE candidate events
are divided into 25 logarithmically even energy bins from 2 GeV to 20
GeV and 240 time bins (6 hours each). The statistical errors in two-dimensional bins  are about
$2\%$ at 5 GeV and $\sim8\%$ at 20 GeV.

Due to the shielding effect of the geomagnetic field, primary GCRs can
only be detected when their rigidities are larger than the cutoff rigidity.
A threshold of $1.2\times$VRC is adopted to minimize such an effect.
Since the VRC varies with locations of the detector, the effective
exposure time thus changes with particle rigidity/energy. In each energy
bin and time bin, the exposure time is calculated via summarizing the
time when the satellite travels in the regions with VRC values smaller
than 1/1.2 of the lower edge of the energy bin, with an additional
subtraction of the dead time of the DAQ system and the time when DAMPE
passes through the SAA region or under the calibration operations.
On average the effective exposure time is about $80\%$ of the total
time for energies above 15 GeV, which gradually decreases to $\sim30\%$
at 2 GeV.

The CRE flux in the $i$th energy bin and $j$th time bin can be calculated as
\begin{equation}
\Phi_{i,j}=\frac{N_{i,j}(1-f_{i,j})}{A_{i}\eta_{i,j}\Delta T_{j}\Delta E_{i}},
\label{eq:fluxfunction}
\end{equation}
where $N_{i,j}$, $f_{i,j}$, and $\eta_{i,j}$ represent the number of
CRE candidates, the fraction of the background contamination, and the
trigger efficiency, $A_{i}$ is the effective acceptance of the detector,
$\Delta T_{j}$ is the live time, and $\Delta E_{i}$ is the width of
the energy bin. In this analysis we focus on the relative fluxes of
CREs in a time scale of two months, and therefore the absolute
efficiences are not crucial.

The dominating systematic uncertainty of the relative flux ratio measurements
comes from the stability of efficiencies. An extensive study of the
efficiency variations have been performed in this analysis. As shown in
\cite{Ambrosi2019}, the major calibration parameters, including the
pedestals, dynode ratios, electronics gains, MIP energies etc.,
are very stable after the temperature correction (with variations
$<1.5$\% per year). Therefore the detection efficiency change within
one month is expected to be very small ($0.13\%$).
On the other hand, due to the radiation damage and electronics aging,
the light yields of the BGO crystals and the gains of photomultipliers
are found to slightly decrease over time. Their impacts on the trigger
threshold of each BGO crystal are carefully calibrated, which show a
decrease of $\sim1.25\%$ per year. We further investigate the impact
of the trigger efficiency from the potential bias of the energy scale
through applying a series of bias factors from $-5\%$ to $+5\%$ with a step of $1\%$ in the MC simulation,
and find that the change of trigger efficiency within 1 month is also small
($0.5\%@2GeV, 0.01\%@20GeV$). The other efficiency is the track efficiency
of the STK. Since we employ track selection at different moments of 2017, we estimate the corresponding stability of the tracking, using proton flight data samples. Our results confirm the high stability of the tracking efficiency, with the overall variations throughout the data taking period less than $0.2\%$ per year.

In summary, the total systematic uncertainty of the relative
detection and selection efficiencies is about $0.53\%$, which
is much smaller than the statistical errors ($2\%\sim8\%$).

\section{Time Profiles of CRE fluxes}
\begin{figure}[htb]
\centering
\includegraphics[width=.90\textwidth]{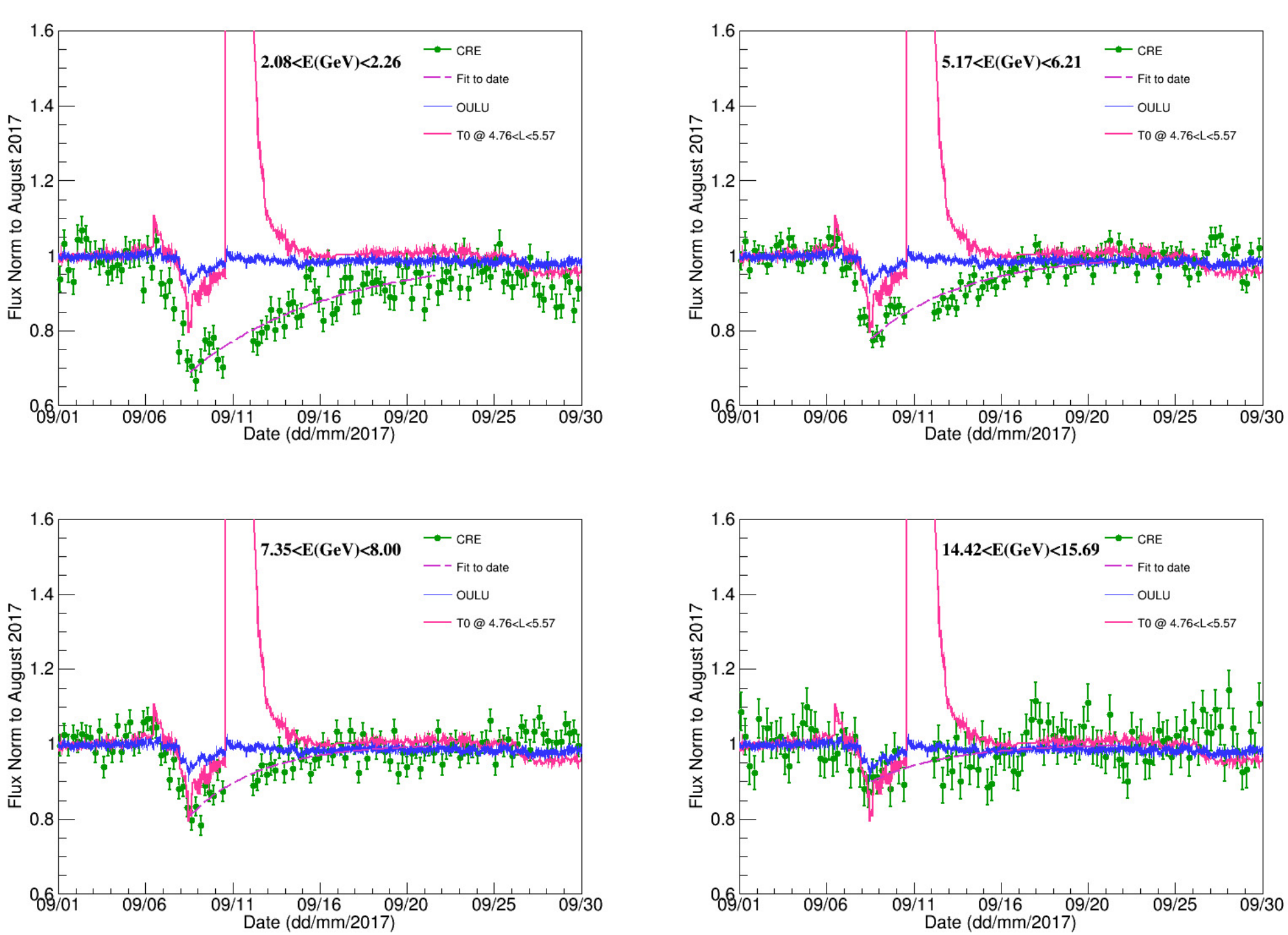}
\caption{Time profiles of CRE fluxes in four energy bins from 00:00
UT of September 1 to 00:00 UT of October 1, 2017. The error bars are summation of
statistical errors and systematic errors. In each plot, the T0 counting rates for
$4.76<L<5.57$ are shown in red and the OULU NM intensities are in blue.
}
\label{fig:CREFD}
\end{figure}

We calculate the CRE fluxes with a time bin width of 6 hours (about 4
orbits for DAMPE). The flux of each bin is normalized to the average
flux of August, 2017. The time profiles of 4 selected energy bins are
shown in Figure \ref{fig:CREFD}. As we have discussed in Sec. 3.1, the
data from 18:00UT of September 9 to 12:00UT of September 11 have been eliminated due to the
strong impact from the energetic particles from the SEP.
The relative intensities averaged over 10 minutes from the OULU
NM\footnote{\url{http://www01.nmdb.eu/nest/}}
are also plotted in Figure \ref{fig:CREFD}.

Compared with the T0 rates, we do not find significant increase in the CRE fluxes during first SEP event. One possible reason is that the data
statistics is limited, and the population of SEP above GeVs is not significant. The NM data do not show the
SEP event either. The physical interpretation of such
differences may need further studies.

\begin{figure}[htb]
\centering
\includegraphics[width=.65\textwidth]{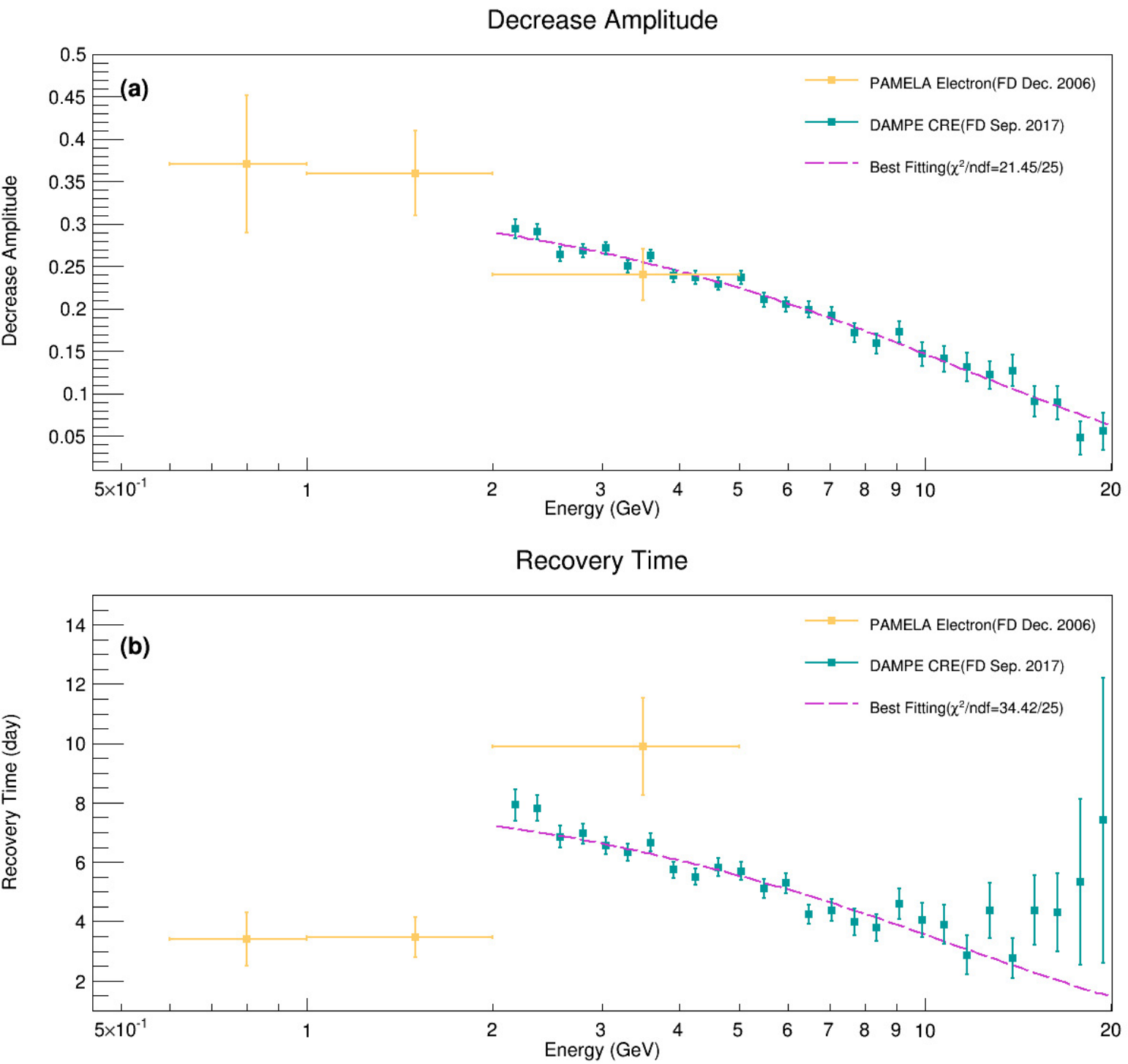}
\caption{The decease amplitude (panel (a)) and recovery time (panel (a))
versus energy. Red solid lines show the best-fits with an exponential
function $p_{0}\exp(-E/p_1)$.}
\label{fig:RTDA}
\end{figure}

Similar with the PAMELA experiment \citep{PAMELAFD2018} we use the following function
\begin{equation}
\frac{\Phi(t)}{\Phi_{\rm ref}}=1-A_{e}\exp\left(-\frac{t-t_0}{\tau}\right),
\label{eq:fittingfunction}
\end{equation}
to describe the amplitude and recovery behavior after the CRE fluxes reach
the valley. In the above equation, $\Phi_{\rm ref}$ is the reference flux
of August, 2017, $A_e$, $\tau$, and $t_0$ represent the decrease
amplitude, recovery time, and time of the minimum flux (or the start
of recovery phase), respectively. For this event, $t_0$ is fixed to be
13:30 UT of September 8, 2017. $A_e$ and $\tau$ parameters are fitted
for each energy bin. The fitting results of the 25 energy bins of $A_e$
and $\tau$ are shown in Figure \ref{fig:RTDA}. Both $A_e$ and $\tau$
clearly show decreases with energy. An exponential decay function,
$p_0\exp(-E/p_1)$, is adopted to fit the energy-dependences of
$A_e$ and $\tau$, resulting in $\chi^2/{\rm ndf}=21.45/25$ and
$\chi^2/{\rm ndf}=34.42/25$, respectively.

The amplitude can be well described by the exponential function.
The fact that the FD amplitudes become smaller for higher energies
can be easily understood as that high energy particles are less
affected by the perturbed interstellar environment by the CME.
Consequently, the energy spectra should also vary with
time when an FD occurs \citep{Belov_2021}, which are expected to be harder
during the FD. A single power-law fit to the energy spectrum between 4 and 20 GeV
shows that the spectral index is about $2.85$ before the FD,
$\sim2.70$ when the fluxes reach the minimum, and becomes $2.85$ after the FD.
The energy-dependence of the recovery time is, however, more
complicated than the amplitude. It is interesting to compare the
results obtained in this study with that by PAMELA.
In \citet{PAMELAFD2018}, the PAMELA experiment gave the energy-dependences
of the recovery time for electrons, protons, and helium nuclei, for
rigidities from 0.6 GV to 10 GV. The qualitative behaviors of recovery
time for these three types of particles are similar, all show an
increase from 0.6 GV to 5 GV, although the specific values for
electrons are different from that for nuclei. At higher energies,
the recovery time for protons and helium nuclei decreases again.
Due to the limited statistics, the decreasing behavior of the
electron recovery time has not been observed. The DAMPE result
shows clearly the decreasing behavior for the CRE recovery time
above 2 GeV, which may indicate that the peaks of the recovery
time for all particles are few GeV. We can further see from
Figure \ref{fig:RTDA} that at high energies, the recovery time tends
to be a constant rather than keeping on decreasing. The recovery
time for protons by PAMELA may also show such a flattening at
high energies \citep{PAMELAFD2018}. But the errors, both for PAMELA
protons and DAMPE CREs, are too large to clearly address this phenomenon.

The ground NMs can also measure the energy-dependences of the FDs,
although the energy resolution is relatively poor and the energy
threshold is relatively high. As illustrated in \citet{Zhao2016},
there are two types of energy-dependences of the recovery time,
one depends strongly on energies and the other remains almost independent
of energies. It has been conjected that the orientations and topologies
of the CMEs may result in such differences \citep{Zhao2016}.
However, a reliable energy determination of the NMs is
difficult. For specific NM station, a typical median energy $E_{M}$
is defined as the value below which cosmic rays contribute to one half
of the counting rate. The median energy can be estimated as a function
of cutoff rigidity $R_c$ as $E_{M}=0.0877R^2_c+0.154R_c+10.12$, where
$E_{M}$ and $R_c$ are in units of GeV and GV \citep{2007AdSpR..40..342J}.
The median energy is $\sim10$ GeV at the polar region and $\sim20$ GeV
at the equator. The full behaviors of the recovery time in a wide energy
range, especially for lower energies, are thus difficult to be revealed
by the NMs. The energy-dependence or independence of the recovery time
shown by NMs may be only a part of the full energy-dependence.
For the event we discuss here, the CME travels towards the Earth.
It should correspond to the Event 2 of \citet{Zhao2016}, and the
recovery time is expected to be a constant. Our results do not support
such an expectation in general. However, the constant behavior may just
be the reflection of the flattening above $\sim10$ GeV as shown in
Figure \ref{fig:RTDA}. A more complete modeling of the recovery time
versus energies is necessary.

\section{Modeling of the CRE FDs}

The Parker's equation has been widely employed to describe the transport
of charged particles in the heliosphere \citep{1965P&SS...13....9P}.
When the heliosphere has been disturbed by CMEs and the associated
shocks, FDs of GCR fluxes can occur. This process can be modelled using
the diffusion barrier model, which employs a moving diffusion barrier
with different diffusion and drift parameters from those in the
interplanetary space \citep{2018ApJ...860..160L}. The Parker's transport
equation together with the diffusion barrier can be solved with the
stochastic differential equation method \citep{1999ApJ...513..409Z,
2011ApJ...735...83S}. We describe the method of the modeling of FDs
in detail in the Appendix.

\begin{figure}[!htb]
\centering
\includegraphics[width=0.45\textwidth]{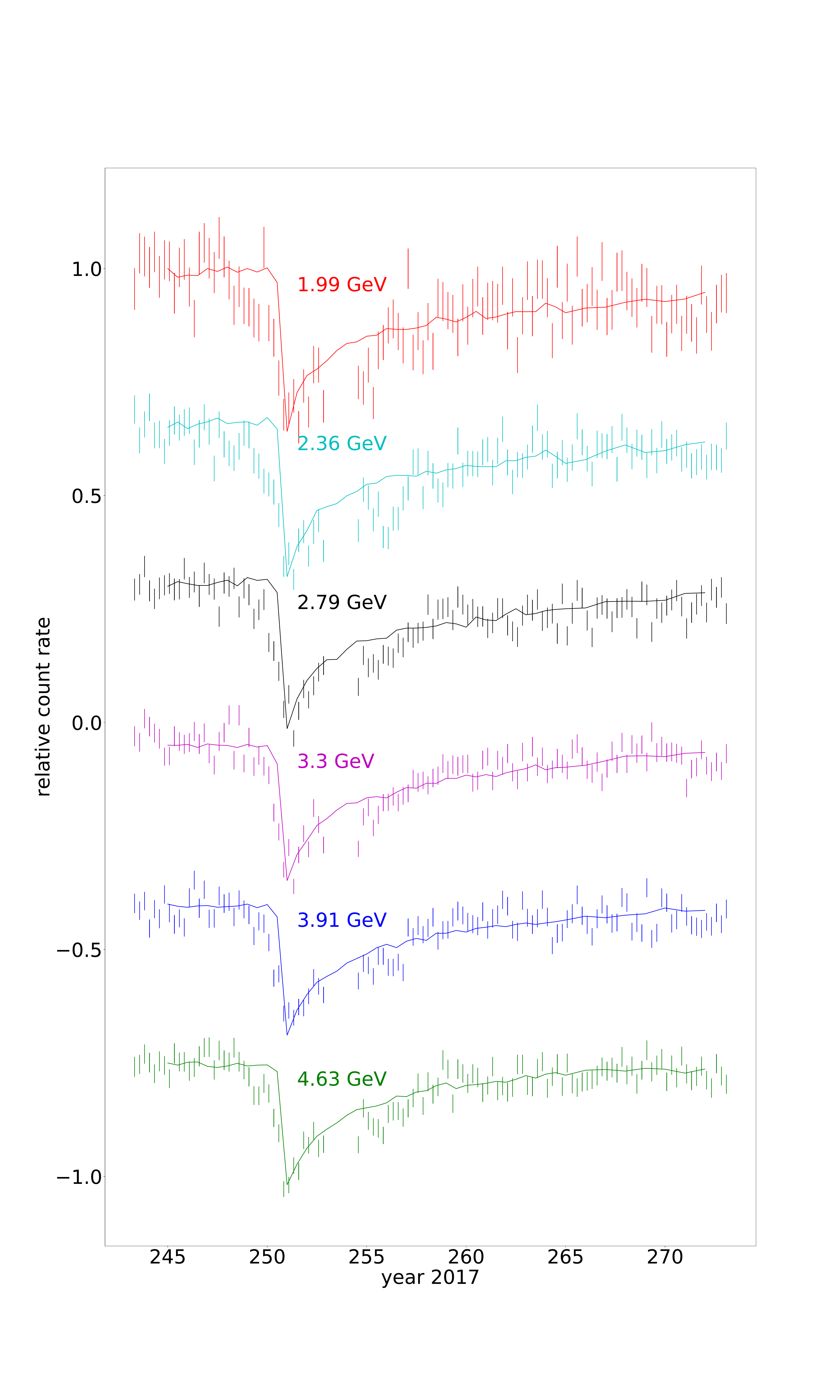}
\includegraphics[width=0.45\textwidth]{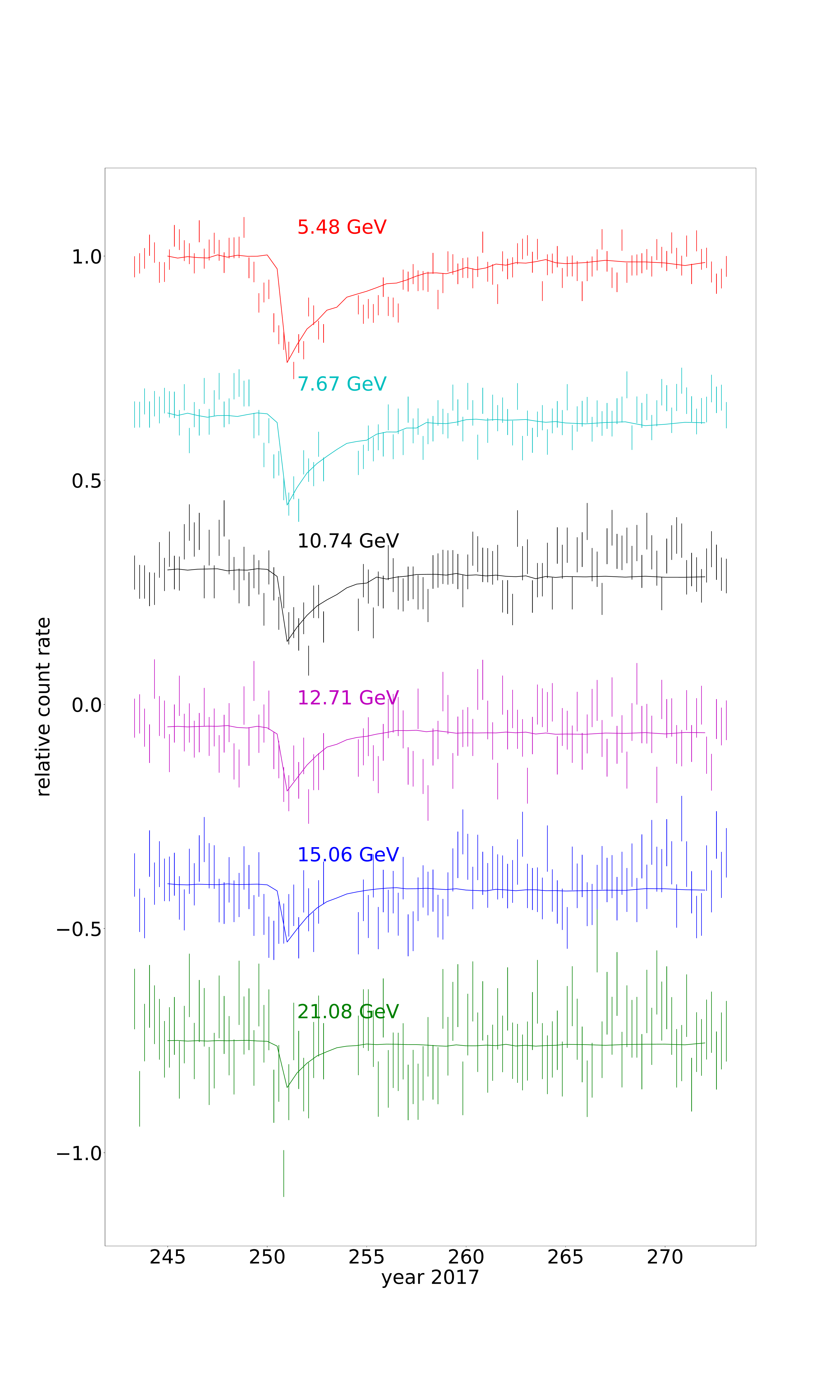}
\caption{Variations of the relative CRE fluxes from the DAMPE
observations (bars) and the FD model (lines). For visualization the
results for different energies are shifted downwards gradually with
respect to the top one in each panel.}
\label{fig:model}
\end{figure}

\begin{figure}[!htb]
\centering
\includegraphics[width=0.45\textwidth]{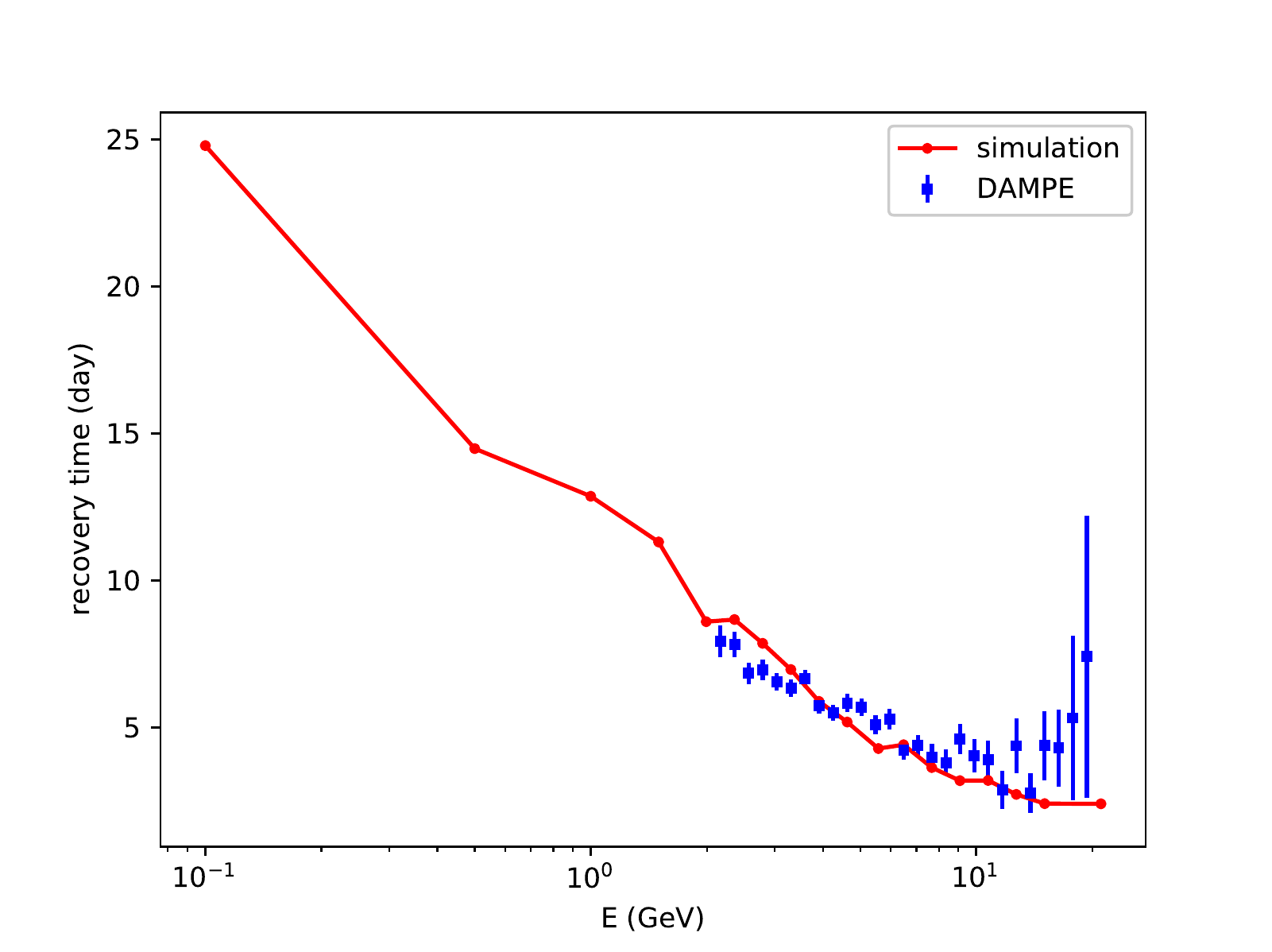}
\caption{The recovery time of the model prediction compared with the
DAMPE data.}
\label{fig:model_t}
\end{figure}

Given proper parameters of the model, the FD behaviors of the DAMPE
data can be reproduced, as shown in Figure \ref{fig:model} for selected
energy bands. We further derive the recovery time at different energies
using Eq.(\ref{eq:fittingfunction}), and give the results in
Figure \ref{fig:model_t}. We can see that the model prediction is in
good agreement with the data in the energy range of $2-20$ GeV.
However, this model may not explain precisely the detailed
behaviors of the data at higher or lower energies. The possible
flattening at high energies, although higher precision of the
measurements is required to reach a conclusion, is not obvious in
the modeling. Also the model predicts longer recovery time at lower
energies, which may not explain the turnover at few GeV as indicated
by the PAMELA and DAMPE data. Further refinement or modification of
the FD modeling is thus necessary to fully account for the data.

\section{Conclusion}

In this work we give a detailed study of the FDs of CREs associated
with the CME occurred in September, 2017 by the DAMPE detector.
A weak SEP event has been revealed through the T0 counting
rate, while no unambiguous signal is shown by on-ground NMs.
Meanwhile the high-precision time evolutions of CRE fluxes with a
6-hour time resolution have been obtained simultaneously.
We study the energy-dependences of the FD amplitudes and the
recovery time, and find that both decrease with energies.
Together with the results of different particle species measured by
PAMELA, it may show that the recovery time of all species of particles
show an increase below a few GeV and a decrease at higher energies.
For energies above $\sim10$ GeV, the recovery time may become
constant, which nevertheless requires more statistics to draw a
firm conclusion. These results would give very important implications
in modeling the propagation of GCRs in the heliosphere when there are
disturbances from CMEs.

This study also illustrates the importance of direct measurements of
FDs of various GCR species. Compared with the usually employed NM data,
the direct measurements have the advantages of excellent particle
identification, very good energy resolution, and much wider energy
range coverage. More comparative studies of various FD events for
different particle species by DAMPE in future are expected to further
shed light on the understanding of the physics of particle
transportation in the interplanetary environment.

\section*{Acknowledgments}
The DAMPE mission is funded by the strategic priority science and
technology projects in space science of Chinese Academy of Sciences.
In China the data analysis is supported by the National Key
Research and Development Program of China (No. 2016YFA0400200), the
National Natural Science Foundation of China (Nos. 11773085, U1738207, 11722328, 41774185,
U1738205, U1738128, 11851305, 11873021, U1631242), the strategic priority science
and technology projects of Chinese Academy of Sciences (No. XDA15051100),
the 100 Talents Program of Chinese Academy of Sciences, the Young
Elite Scientists Sponsorship Program by CAST (No. YESS20160196),
and the Program for Innovative Talents and Entrepreneur in Jiangsu.
In Europe the activities and the data analysis are supported by the
Swiss National Science Foundation (SNSF), Switzerland; the National
Institute for Nuclear Physics (INFN), Italy.


\bibliography{DAMPE_CRE_FD_APJ}{}

\bibliographystyle{aasjournal}

\clearpage

\appendix

\section{The Parker's transport equation}

The Parker's transport equation is written as \citep{1965P&SS...13....9P}
\begin{eqnarray}
\label{eq:parker}
\frac{\partial f}{\partial t} & = & -~\left( \vec{V}_{\rm sw}+
\left\langle\vec{v}_d\right\rangle \right) \cdot \nabla f \\
& + &\nabla \cdot \left(K^{(s)} \cdot \nabla f\right) \nonumber \\
& + &~\frac{1}{3} \left(\nabla \cdot \vec{V}_{\rm sw}\right)
\frac{\partial f}{ \partial \ln p},
\end{eqnarray}
where $f(\vec{r},p,t)$ is the particle distribution function,
$p$ is the momentum, $\vec {V}_{\rm sw}$ is the latitude-dependent
solar wind speed \citep{2015ApJ...810..141P,2019ApJ...878....6L},
$\left \langle \vec v_d \right \rangle$ is the pitch-angle averaged
drift velocity \citep{1977ApJ...213..861J,1989ApJ...339..501B,1974ApJ...192..535F,2015AdSpR..55..390E} with the following form:
\begin{equation}\label{eq:drift}
\left \langle \vec v_d \right \rangle =\frac{pv}{3q}
\nabla \times \frac{\vec B}{B^{2}},
\end{equation}
where $q$ and $v$ are the charge and velocity of the GCR particle,
and $\vec{B}$ is the magnetic field inside the heliosphere.
The last term of Eq.(\ref{eq:parker}) is the adiabatic energy loss.
$K^{(s)}$ is the symmetric diffusion tensor, given in the IMF aligned
coordinates by
\begin{equation}
K^{(s)} = \left( \begin{array}{ccc} \kappa_{\parallel} &0 &0\\0& \kappa_{\perp \theta}& 0\\ 0& 0& \kappa_{\perp r}\\
\end{array} \right),
\end{equation}
where $\kappa_{\parallel}$ is the parallel diffusion coefficient,
and $\kappa_{\perp r}$, $\kappa_{\perp\theta}$ are two perpendicular
diffusion coefficients in the radial and latitudinal directions.
In our simulation, the diffusion coefficients are expressed as
\citep{2019ApJ...878....6L}
\begin{eqnarray}
\kappa_{\parallel}  =  \kappa_{\parallel 0} \beta \frac{B_0}{B}
\left(\frac{P}{P_k}\right)^a\left[1+\left(\frac{P}{P_k}\right)
^{\frac{b-a}{c}}\right]^c,\\
\kappa_{\perp r ,\perp \theta }  =  (\kappa_{\perp r 0 ,\perp \theta 0})
\beta \frac{B_0}{B}\left(\frac{P}{P_k}\right)^a\left[1+\left(\frac{P}{P_k}
\right)^{\frac{b-a}{c}}\right]^c.
\end{eqnarray}
Here, $\beta$ is the particle velocity in unit of the light speed,
$B$ is the magnitude of the local IMF with $B_0= 1$~nT, $P$ is the
rigidity, $\kappa_{\parallel 0}$, $\kappa_{\perp r 0 }$, and
$\kappa_{\perp \theta 0 }$ are constants, $P_k$, $a$, $b$, $c$ are
free parameters which define the rigidity dependence of these
diffusion coefficients.

We use the standard Parker field to describe the IMF
\citep{1958ApJ...128..664P}:
\begin{equation}
\vec{B}(r,\theta)=\frac{AB_\oplus}{r^2}\left( \boldsymbol{e_r} -
\frac{r\Omega \sin \theta}{V_{\rm sm}} \boldsymbol {e_\theta}\right)
\left[ 1-2H(\theta-\theta_{\rm cs})\right ],
\end{equation}
where $B_\oplus$ is the reference value at the earth position which
we take as the mean value of the last 13 months before the solar flare
event, $A = \pm 1/{\sqrt {\left( 1+\frac{\Omega^2}{\vec V_{\rm sw}^2}\right)}}$
with the sign determining the polarity of the IMF, $\Omega$ is the
angular velocity of the Sun, $H(\theta-\theta_{\rm cs})$ is the Heaviside
function and $\theta_{\rm cs}$ determines the polar extent of the
Heliospheric Current Sheet (HCS). We model the HCS by the following
analytical expression \citep{1983ApJ...265..573K},
\begin{equation}
\cot\theta_{\rm cs}=-\tan \alpha ~ \sin\phi^*.
\end{equation}
Here, $\alpha$ is the HCS tilt angle which we also adopt as the mean
value of the last 13 months, $\phi^*=\phi+\frac{r\Omega}{V_{\rm sw}}$
with $\phi$ being the longitude angle of the current sheet surface.
We do not consider the phase of the co-rotating HCS.

The the magnitude of drift in the HCS can be specified as
\citep{1989ApJ...339..501B}
\begin{equation}
v_{\rm HCS} = \frac{v}{6} \times \frac{4R_g}{r}\delta(\theta-\theta_{\rm cs}),
\end{equation}
where $R_g$ is the gyroradius of the GCR particle and $v$ is the particle's
speed. Following \citet{2013JGRA..118.7517L} and \citet{2017ApJ...839...53L},
we replace $\delta (\theta-\theta_{\rm cs})$ with $r/(4R_g)$ within the
distance of $(-2R_g, 2R_g)$ to the HCS. By defining $\psi$ as the angle
between the local current sheet normal direction $\boldsymbol{n}$ and
the $-\boldsymbol{e_\theta}$ direction, the drift vector of the current
sheet can be written as
\begin{equation}
\vec{V}_{\rm cs}=\frac{v}{6}\left(\boldsymbol{e_r}\sin {\Phi} \cos \psi
+ \boldsymbol{e_\theta}\sin\psi+\boldsymbol{e_\phi}\cos\Phi \cos\psi\right),
\end{equation}
with $\Phi$ being the IMF winding (spiral) angle.

\section{The diffusion barrier model}
In our model, we use a 3D geometry profile to describe the propagating
diffusion barrier. The diffusion and drift coefficients inside the
barrier are expressed as
\begin{equation}
\kappa'_{\parallel,\perp,T}=\frac{\kappa_{\parallel,\perp,T}}{1+\rho h(\theta)f(r)g(\phi)}.
\end{equation}
In the above equation, $\kappa'_{\parallel,\perp}$ ($\kappa'_T$) are
the diffusion (drift) coefficients inside the barrier, $\rho$ is a
constant determining the reduction level of the diffusion and drift
coefficients, $h(\theta)$, $g(\phi)$, and $f(r)$ describe the geometry
of the diffusion barrier
\begin{eqnarray}
h(\theta)&=&e^{-\left( \frac {\theta-\theta_0}{\theta_{\rm br}} \right)^{10}},\\
g(\phi)&=&e^{-\left( \frac {\phi}{\phi_{\rm br}} \right)^{10}},\\
f(r)&=& \left\{\begin{array}{lc}
    1-\frac{r-r_{\rm cen}}{r_{\rm a}}, & r_{\rm cen}<r<r_{\rm sh}\\
    \frac{r-r_{\rm end}}{r_{\rm b}}, & r_{\rm end}<r\le r_{\rm cen}\\
    0, & {\rm otherwise}
\end{array}\right..
\end{eqnarray}
Here, $\theta_0=\pi/2$ means that diffusion barrier hits the earth
head on, the parameters $\theta_{\rm br}$ and $\phi_{\rm br}$ denote
the extensions of the diffusion barrier along the $\boldsymbol{e_\theta}$
and $\boldsymbol{e_\phi}$ directions, $r_{\rm sh}$, $r_{\rm cen}$,
and $r_{\rm end}$ are the radial distances of the front, center,
and end locations of the barrier, $r_{\rm a}=r_{\rm sh}-r_{\rm cen}$,
$r_{\rm b} =r_{\rm cen} -r_{\rm end}$ are widths of the leading and
trailing paths of the barrier. In our model, we assume $r_{\rm a}=0.5$ AU,
$r_{\rm b}=0.3$ AU, $\theta_{\rm br}=60^\circ$, $\phi_{\rm br}=180^\circ$.
The velocity of the diffusion barrier is adopted to be 1500 km/s, as given
in the CME list\footnote{https://cdaw.gsfc.nasa.gov/CME\_list/}.

\section{The stochastic differential equation method}
Based on the Ito's formula, the Parker's equation can be solved
numerically with the stochastic differential equation (SDE) technique
\citep{1999ApJ...513..409Z,2011ApJ...735...83S}, with
\begin{eqnarray}
d \vec{X} = \left(\nabla \cdot K^{(s)} - \vec{V_{\rm sw}} -
\left \langle \vec v_d \right \rangle\right)ds +
\sum^3_{\sigma=1}\vec{\alpha}_{\sigma}dW_{\sigma}(s),\\
dp = \frac{p}{3}(\nabla \cdot \vec{V}_{\rm sw}\,ds),
\end{eqnarray}
where $dW_{\sigma}(s)$ is the Wiener process, which can be numerically
generated by the Gaussian random number, and
$\sum_\sigma{\alpha}_{\sigma}^{\mu}\alpha_{\sigma}^{\nu}=2\kappa ^{\mu\nu}$.
The SDE method gives the time-dependent solution of Eq.~(\ref{eq:parker})
as \citep{1999ApJ...513..409Z,2019ApJ...878....6L}
\begin{equation}
f(t,\vec X,p)=\left\langle f_b^i(t-\chi_i,\vec X_i,p_i) \right \rangle.
\end{equation}
$\left\langle f_b^i(t-\chi_i,\vec X_i,p_i) \right \rangle$ is the
mean value for the pseudo-particles reaching the boundary
$(\vec X_i,p_i)$ at the first exit time $\chi_i$. In our calculation,
the local interstellar spectrum of electrons are adopted from
\citet{2021APh...12402495Z}. The small fraction of positrons has
been neglected.

\end{document}